\documentclass[%showpacs, showkeys,
12pt,
preprint,preprintnumbers,nofootinbib,
groupedaddress,superscriptaddress,amsmath,amssymb]{revtex4}
\usepackage{graphicx}% Include figure files
\usepackage{dcolumn}% Align table columns on decimal point
\usepackage{bm}% bold math
\usepackage{amssymb}
\usepackage{amsmath}
\usepackage{epsfig}    
\usepackage{color}
\usepackage{slashed}
\usepackage{hhline}
\def\be{\begin{equation}}
\def\ee{\end{equation}}
\newcommand{\bea}{\begin{eqnarray}}
\newcommand{\eea}{\end{eqnarray}}
\newcommand{\nn}{\nonumber}

\numberwithin{equation}{section}

\begin{document}

{\begin{flushright}{KIAS-P17039}
\end{flushright}}

%%%%%%%%%
\title{Gauged Lepton Symmetry Model}
%\preprint{KIAS-P14078}
%

\author{Takaaki Nomura}
\email{nomura@kias.re.kr}
\affiliation{School of Physics, KIAS, Seoul 02455, Korea}

\author{Hiroshi Okada}
\email{macokada3hiroshi@cts.nthu.edu.tw}
\affiliation{Physics Division, National Center for Theoretical Sciences, Hsinchu, Taiwan 300}

\date{\today}

\begin{abstract}
We  study a model with $U(1)$ gauged lepton number symmetry in which the active neutrino masses are generated at two-loop level through the spontaneous breaking of the lepton number  symmetry.  To cancel gauge anomalies some exotic leptons are introduced that also play an important role in generating neutrino masses and the lightest neutral one can be a dark matter candidate. We discuss a phenomenology of the neutral fermion sector including dark matter, lepton flavor violations, and an extra neutral gauge boson $Z'$ that can be tested by $e^+ e^-$ collision at the International Linear Collider (ILC).
\end{abstract}
\maketitle
\newpage

\section{Introduction}

The lepton number is conserved as a global symmetry in the framework of the standard model (SM) accidentally, which
motivates us to promote its symmetry to the new gauge symmetry.
A gauged $U(1)$ lepton symmetry extended model has been done by the group in ref.~\cite{Aranda:2014zta}, in which they mainly discuss the electroweak phase transition. 
Also a gauged $SU(2)$ lepton symmetry extended model has been done by the group in ref.~\cite{Fornal:2017owa}, in which an asymmetric dark matter (DM) is mainly discussed.
In these models, some exotic leptons are introduced to cancel gauge anomalies associated with gauged lepton number symmetry.

%The lepton number is always violated when the active neutrinos are Majorana fermions.Supposing a canonical seesaw at the tree level, for example, heavy Majorana fermions are the origin of violating the lepton number. 
%Thus once the heavy fermions generate their masses after the spontaneous breaking of lepton number symmetry,  the active neutrino masses can be connected to the spontaneous breaking of lepton number symmetry.

Even when active neutrino mass are radiatively induced,
the lepton number is always violated in case of Majorana fermions. 
%whenever a Majorana neutrino mass term is generated through loop effect. 
Thus the active neutrino masses can be connected to spontaneous breaking of lepton number symmetry when it is gauged.
It is therefore interesting to construct a radiative neutrino mass model with gauged lepton number symmetry and to discuss resulting phenomenology.

In this paper, we construct a lepton model with a gauged $U(1)$ lepton number symmetry, introducing exotic leptons
in order to cancel the anomalies out among lepton sectors.
Then the active neutrino masses are induced at two-loop level where the exotic leptons play a role in propagating inside a loop diagram.~\footnote{Several neutrino mass models at two-loop level are found in \cite{2-lp-zB, Babu:2002uu, AristizabalSierra:2006gb, Nebot:2007bc, Schmidt:2014zoa, Herrero-Garcia:2014hfa, Long:2014fja, VanVien:2014apa, Aoki:2010ib, Lindner:2011it, Baek:2012ub, Aoki:2013gzs, Kajiyama:2013zla, Kajiyama:2013rla, Baek:2013fsa, Okada:2014vla, Okada:2014qsa, Okada:2015nga, Geng:2015sza, Kashiwase:2015pra, Aoki:2014cja, Baek:2014awa, Okada:2015nca, Sierra:2014rxa, Nomura:2016rjf, Nomura:2016run, Bonilla:2016diq, Kohda:2012sr, Dasgupta:2013cwa, Nomura:2016ask, Nomura:2016pgg, Liu:2016mpf, Nomura:2016dnf, Simoes:2017kqb, Baek:2017qos, Ho:2017fte}.}
Such extra fermions are also assumed to be odd under $Z_2$ symmetry and the lightest neutral component can be a good DM candidate.
We discuss the neutrino mass, lepton flavor violation (LFV), muon $g-2$, relic density of dark matter, and collider physics associated with $Z'$ boson.

This paper is organized as follows.
In Sec.~II, we show our model, %to introduce exotic fermions and bosons with some additional symmetries,  
and formulate the neutral fermion sector including active neutrinos, (vector-gauge) boson sector, lepton sector, dark matter sector, and discuss collider physics through $Z'$ boson at ILC .
Finally We conclude and discuss in Sec.~III.
%\newpage

%%%%%%%%%%%%%%%%%%%%%%%%%%%%%%%%%%%%%
%\section{The Model}
%\subsection{Model setup}

 \begin{widetext}
\begin{center} 
\begin{table}[b]%[tbc]
%\begin{tiny}
\begin{tabular}{|c||c|c|c||c|c|c||c|c|c|||c|c|c|c|}\hline\hline  
%&\multicolumn{5}{c||}{SM leptons} & \multicolumn{3}{c|}{Exotic fermions} \\\hline
 & ~$L_{L_a}$~ & ~$e_{R_a}$~ & ~$N_{R_a}$~ &~$L''_R$~ & ~$e''_L$~ & ~$N''_{L}$~ &~$L'_L$~ & ~$e'_R$~ & ~$N'_{R}$~ 
 & ~$H$~ & ~$\varphi$~  & ~$S_1$~ & ~$S_2$~
\\\hline 
%$SU(3)_C$ & $\bm{3}$  & $\bm{3}$  & $\bm{3}$  & $\bm{1}$  & $\bm{1}$  & $\bm{1}$  & $\bm{1}$  & $\bm{1}$  \\\hline 
 %%%
 $SU(2)_L$ & $\bm{2}$  & $\bm{1}$  & $\bm{1}$ & $\bm{2}$ & $\bm{1}$  & $\bm{1}$& $\bm{2}$ & $\bm{1}$  & $\bm{1}$ & $\bm{2}$ 
 & $\bm{1}$  & $\bm{1}$ & $\bm{1}$   \\\hline 
 %%%
$U(1)_Y$ & $-\frac12$ & $-1$  & $0$ & $-\frac12$  & $-1$ & $0$& $-\frac12$  & $-1$ & $0$ & $0$ & $0$& $0$ & $0$    \\\hline
 %%%
 $U(1)_{L}$ & $1$ & $1$  & $1$ & $2$  & $2$ & $2$  & $-1$  & $-1$   & $-1$   & $0$   & $2$    & $1$ & $1$ \\\hline
 $Z_2$ & $+$ & $+$ & $-$ & $-$ & $-$ & $-$ & $-$ & $-$ & $-$ & $+$ & $+$ & $+$ & $-$ \\ \hline
 %%%
\end{tabular}
\caption{Field contents of fermions and bosons
and their charge assignments under $SU(2)_L\times U(1)_Y\times U(1)_{L}$, where $a(=1-3)$ is flavor indices.}
\label{tab:1}
% \end{tiny}
\end{table}
\end{center}
\end{widetext}

\section{Model setup and phenomenologies}
In this section, we construct our model and discuss resulting phenomenologies.
First of all, we impose an additional $U(1)_{L}$ gauge symmetry and discrete $Z_2$ symmetry with three right-handed neutral fermions $N_{R_a}(a=1-3)$ which are charged under the $U(1)_L $ and odd under the $Z_2$.
In addition, we also introduce the two sets of $Z_2$ odd families $(L''_R,e''_L,N''_L)$ and $(L'_L,e'_R,N'_R)$ with the same charges under the SM gauge groups, but with  $(2,-1)$ charges under $U(1)_L$ symmetry respectively.~\footnote{In general, $-\ell''+\ell'=-3$ provides the anomaly free theory, where $\ell''$ is the $U(1)_L$ charge of $(L''_R,e''_L,N''_L)$, and  $\ell'$ is the one of $(L'_L,e'_R,N'_R)$}
 %  which are chiral-flipped fields against $(L_L,e_R,N_R)$ and the number of three charges correspond to the number of families.~\footnote{Thus we have a possibility to introduce three families for one charges under the $U(1)_L$ symmetry for $(L'_R,e'_L,N'_L)$. However it conflicts to reproduce the neutrino oscillation data due to the absence of massive active neutrinos.}

Thus $[U(1)_L]^2\times U(1)_Y$, $[U(1)_L]^3$, and $U(1)_L$ gauge anomalies are separately cancel out each in $(L_L,e_R,N_R)$ and $(L''_R,e''_L,N''_L,L'_L,e'_R,N'_R)$ sectors. 
{On the other hand, $[U(1)_Y]^2\times U(1)_L$ and $[SU(2)]^2 \times U(1)_L$ nontrivially cancel the anomaly out between $(L_L,e_R,N_R)$ and $(L''_R,e''_L,N''_L,L'_L,e'_R,N'_R)$; each sector provides the values of $\sum Q_Y^2 Q_L (\sum {\rm Tr}[\sigma^a \sigma^b Q_L])$ as $3/2(6)$ and $-3/2(-6)$ respectively where $Q_{Y(L)}$ denotes hypercharge(lepton number) and $\sigma^a$ is the Pauli matrix.}
%%%%%%%%% %%%%%%%%% %%%%%%%%%
In the boson sector, we introduce an  isospin singlet scalar field $\varphi$ to break the  $U(1)_{L}$ symmetry spontaneously.
For neutrino mass generation, we also add $U(1)_{L}$ charged singlet scalar fields $S_1$ and $S_2$ which are respectively $Z_2$ even and odd. 
Field contents and their assignments for fermions and bosons are respectively summarized in Table~\ref{tab:1}.

Under these symmetries, the renormalizable Lagrangian for lepton sector and Higgs potential are respectively given by 
\begin{align}
%\label{eq:tri-y}
-{\cal L}_{} & \supset
M'_{1a} \bar N'^C_R N_{R_a} + y_{\ell_a}\bar L_{L_a} e_{R_a} H %+ (y_\nu)_{ab}\bar L_{L_a} \tilde H N_{R_b}
 +y_{N_{a}} \bar N^C_{R_a}  N_{R_a} \varphi^*\nn\\
&+ y''_e\bar L''_R H e''_L + y''_\nu \bar L''_R \tilde H N''_L + y'_e\bar L'_L H e'_R + y'_\nu \bar L'_L \tilde H N'_R  + y_{N'} \bar N'^C_R N'_R\varphi  
\nn \\
%%%
&+ (y_{L''L})_{1b} \bar L''_R L_{L_b} S_2 + (y_{e''e})_{1b} \bar e''_L e_{R_b} S_2 
+ (y_{N''N})_{1a} \bar N''_L N_{R_a} S_1 
+{\rm c.c.}   \label{eq:nontri-y} ,\\
%%% %%%
V&= \mu_H^2 H^\dag H  + \mu^2_{\varphi}\varphi^* \varphi  +\mu_{S_1}^2 S_1^*S_1 +\mu_{S_2}^2 S_2^*S_2 
+\mu_1 [S_1^2\varphi^*+S_1^{*2}\varphi] + \mu_2 [S_2^2\varphi^*+S_2^{*2}\varphi], \nn \\
& +\lambda_H (H^\dag H)^2 + \lambda_{\varphi}(\varphi^* \varphi)^2 + \lambda_{S_1}(S_1^* S_1)^2 + \lambda_{S_2}(S_2^* S_2)^2 + \lambda_{H\varphi} (H^\dag H)(\varphi^* \varphi) \nn\\
&
+ \lambda_{HS_1} (H^\dag H)(S_1^* S_1) + \lambda_{HS_2} (H^\dag H)(S_2^* S_2) + \lambda_{\varphi S_1}(\varphi^* \varphi)(S_1^* S_1) 
+ \lambda_{\varphi S_2}(\varphi^* \varphi)(S_2^* S_2) \nn \\
& + \lambda_{S_1 S_2}(S_1^* S_1)(S_2^* S_2) + \tilde \lambda_{S_1 S_2}[(S_1^* S_1^*)(S_2 S_2) + h.c.], 
\label{eq:lag-lep}
\end{align}
where $L''_R\equiv [N''_R,e''_R]^T$, $L'_L\equiv [N'_L,e'_L]^T$, $\tilde H \equiv (i \sigma_2) H^*$ with $\sigma_2$ being the second Pauli matrix, and $(a,b)$ runs over $1$ to $3$.
%%%%%%%%%

\subsection{Scalar sector and $Z'$ boson}

{\it Scalar sector}:　
The scalar fields are parameterized as 
\begin{align}
%\begin{tiny}
&H =\left[\begin{array}{c}
w^+\\
\frac{v + h +i z}{\sqrt2}
\end{array}\right],\quad 
%%%
\varphi=
\frac{v'+\sigma + iz'_{\varphi}}{\sqrt2},\quad 
%%%
S_{1,2}=
\frac{s_{R_{1,2}} + i s_{I_{1,2}}}{\sqrt2},
\label{component}
%\end{tiny}
\end{align}
where $w^+$, $z$, and $z'_\varphi$ are absorbed by the SM and $U(1)_L$  gauge bosons $W^+$, $Z$, and $Z'$.
%%%
Inserting tadpole conditions, the CP even mass matrix in basis of $(\sigma, h)$ with nonzero VEVs is diagonalized by 
\begin{align}
O M_R^2 O^T
&\equiv
\left[\begin{array}{ccc}
2 v'^2 \lambda_{\varphi}  &  v v' \lambda_{H\varphi} \\ 
 v v' \lambda_{H\varphi}  & 2 v^2 \lambda_{H} \\ 
\end{array}\right]
= 
\left[\begin{array}{ccc}
v'^2 \lambda_{\varphi} + v^2 \lambda_{H} - D &  0 \\ 
0 &  v'^2 \lambda_{\varphi} + v^2 \lambda_{H} + D \\ 
\end{array}\right]\equiv 
\left[\begin{array}{ccc}
m_{H_1}  &  0 \\ 
0 & m_{H_2} \\ 
\end{array}\right],\\
%%%
O
&\equiv
\left[\begin{array}{ccc}
-c_\alpha &  s_\alpha \\ 
s_\alpha & c_\alpha \\ 
\end{array}\right],\quad s_\alpha =\frac{2 v v'\lambda_{H\varphi}}{m_{H_1}^2-m_{H_2}^2},
\end{align}
where $D\equiv \sqrt{(v'^2 \lambda_{\varphi} - v^2 \lambda_{H})^2 + (vv'\lambda_{H\varphi})^2} $, and $s_{\alpha}(c_{\alpha})$ is the short hand notation of $\sin\alpha(\cos\alpha)$.
  
  %%%%%%%%%%
  {
  {\it $Z'$ boson}: After spontaneous $U(1)_L$ gauge symmetry breaking, we have $Z'$ boson that couples not to quarks but leptons.
  The mass of $Z'$ is given by $m_{Z'} = 4 g' v'$ where $g'$ is the $U(1)_L$ gauge coupling.
  Since our $Z'$ universally couples to SM leptons the LEP experiment provides the strongest constraints on the gauge coupling and $Z'$ mass.
  Assuming $m_{Z'} \gtrsim 200$ GeV, the LEP constraint is applied to the effective Lagrangian 
\begin{equation}
L_{eff} = \frac{1}{1+\delta_{e \ell}} \frac{g'^2}{m_{Z'}^2} (\bar e \gamma^\mu e)( \bar \ell \gamma_\mu \ell)
\end{equation}
where $\ell = e$, $\mu$ and  $\tau$.
We then obtain following constraint from the analysis of data by measurement at LEP~\cite{Schael:2013ita}: 
%which  tells us the following restriction:
\begin{align}
\frac{m_{Z'}}{g'} \gtrsim 7.0\ {\rm TeV}.\label{eq:lep}
\end{align}
We will take into account this constraint in the analysis of DM relic density and collider physics below.
 }

  %%%%%%%%%%%
  
  \subsection{Fermion masses}
  %\subsection{Fermion Sector}
{\it Charged-leptons}:
{The SM charged lepton mass eigenvalues are given by $[m_e,m_\mu,m_\tau]\equiv [y_{\ell_1} v/\sqrt2,y_{\ell_2} v/\sqrt2, y_{\ell_3} v/\sqrt2]$ and extra singly charged leptons are given by $m_{e''}\equiv y''_e v/\sqrt2$ and  $m_{e'}\equiv y'_e v/\sqrt2$ for our minimal filed contents.
 Notice here that the masses of SM charged-leptons does not mix the others.
We also note that diphoton decay branching ratio of SM Higgs, $BR(h_{SM} \to \gamma \gamma)$, is modified by heavy charged lepton loop effects 
 due to large Yukawa coupling constants $y'_{e}$ and $y''_e$.
 Since the branching ratio is strongly constrained by the Higgs measurements at LHC we need to suppress new physics contributions.
 One way is to introduce charged singlet scalar $S^\pm$ which has sizable couplings to the SM Higgs, and the charged scalar loop can cancel the contribution from heavy charged lepton loop. 
 Another way is to obtain heavy charged lepton masses via VEV of $U(1)_L$ charged scalar singlet as discussed in ref.~\cite{Aranda:2014zta}.
 These resolutions do not affect our mechanism of neutrino mass generation and other phenomenologies which will be analyzed below.
 We thus abbreviate detailed analysis of the issue in this paper, and masses of $e'$ and $e''$ are simply parametrized as $M_{e'}$ and $M_{e''}$ respectively.}
%%%

{\it  Exotic neutral fermions}:
We have two mass matrices of neutral fermions in basis of  $\Psi\equiv [N''_R,N''^C_L]^T$ and  $\Psi'\equiv[N_{R_a},N'_R]^T$,
and they are given by
\begin{align}
M_{\Psi}
&\equiv
\left[\begin{array}{ccc}
0 &  m''_D \\ 
 m''_D & \delta m \\ 
\end{array}\right], \quad
%%%
M_{\Psi'}
\equiv
\left[\begin{array}{cc}
(M_N)_{3\times3} &  (M'^T)_{3\times1} \\ 
 (M')_{1\times3} & m_{N'} \\ 
\end{array}\right]
=
\left[\begin{array}{cccc}
m_{N_1} &0 & 0 &  M'_{11} \\ 
0 &m_{N_2} & 0 & M'_{12} \\ 
0  & 0 &m_{N_3}& M'_{13} \\ 
 M'_{11}  &  M'_{12}  & M'_{13}  & m_{N'} \\ 
\end{array}\right],
 \end{align}
where $m''_D\equiv y''_Dv/\sqrt2$, $M_N\equiv Y_{N_{a}}v'/\sqrt2$, $m_{N'}\equiv y_{N'}v'/\sqrt2$, and  $\delta m$ is given at the one-loop level as shown below. Then the mass eigenstates and their mixing are respectively defined by $D_{\psi}=V M_{\Psi} V^T$ and $D_{\psi'}=U M_{\Psi'} U^T$, and 
\begin{align} 
\left[\begin{array}{c}
N''_R  \\ 
N''^C_L \\ 
\end{array}\right]=V^T
\left[\begin{array}{c}
\psi_{1R}  \\ 
\psi_{2L}^C\\ 
\end{array}\right],\quad
%%%
\left[\begin{array}{c}
N_{R_{1\sim3}}  \\ 
N'_R \\ 
\end{array}\right]=U^T
\left[\begin{array}{c}
\psi'_{R_{1\sim3}}  \\ 
\psi'_{R_4}\\ 
\end{array}\right],
   \end{align}
where $V$ and $U$ are respectively two by two and four by four unitary mixing matrices.
%%%
The form of $\delta m$, which is given at the one-loop level, is found to be
 \begin{align}
& \delta m=-\frac{2}{(4\pi)^2}\sum_{\alpha=1}^4(Y_{N''N})_{1\alpha} D_{\psi'_\alpha}(Y_{N''N}^T)_{\alpha1} F_I^\alpha(r'_{R_1},r'_{I_1}),\\
 %%%
& F_I^\alpha(r_1,r_2)=\frac{r_1r_2\ln\left[\frac{r_2}{r_1}\right] + r_1 \ln\left[{r_1}\right] - r_2 \ln\left[{r_2}\right]}
{(r_1-1)(r_2-1)},\label{eq:lpfunc}
 \end{align}
where $(Y_{N''N})_{1\alpha}\equiv \sum_{a=1}^3(y_{N''N})_{1a}(U^T)_{a\alpha}/\sqrt2$ and $r'_{R_1(I_1)}\equiv[m_{S_{R_1}}(m_{S_{I_1}} )/M_{\psi'_\alpha}]^2$.
%%% %%%

%%%%%%%%%%%%%%%%%%%
\begin{figure}[t]
\begin{center}
\includegraphics[width=80mm]{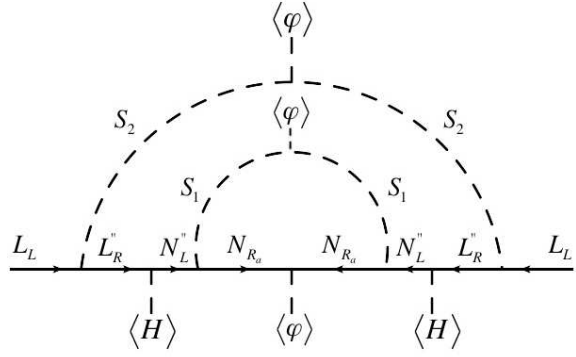} 
\caption{The two loop diagram to generate neutrino masses. } 
  \label{fig:neut}
\end{center}\end{figure}
%%%%%%%%%%%%%%%%%%%

{\it Active neutrinos}:
The neutrino mass matrix is induced at the two-loop level in fig.~\ref{fig:neut}, which
 is given by
\begin{align}
&(m_\nu)_{ab}\approx 
-\frac{2}{(4\pi)^2}\sum_{\alpha=1}^2(Y_{\nu}^T)_{a\alpha} D_{\psi_\alpha}(Y_{\nu})_{\alpha b} F_I^\alpha(r_{R_2},r_{I_2}),
\end{align}
where $(Y_{\nu})_{\alpha a}\equiv  (y_{L''L})_{1a}(V^*)_{\alpha1}/\sqrt2$ is the two by three matrix and $r_{R_2(I_2)}\equiv[m_{S_{R_2}}(m_{S_{I_2}})/M_{\psi_\alpha}]^2$.
%%%
Since one diagonalizes neutrino mass matrix as $D_\nu\approx V_{MNS}m_\nu V_{MNS}^T$, 
we can rewrite Yukawa coupling in terms of neutrino oscillation data and some parameters as:
\begin{align}
(Y_\nu)_{2\times3} =(R^{-1/2} O \sqrt{D_\nu} V^*_{MNS})_{2\times3},
\end{align}
where $ R _\alpha  \equiv D_{\psi_\alpha}F_I^\alpha(r_{R_2},r_{I_2})$, and $O$ is an arbitral two by three orthogonal matrix: $OO^T=1_{2\times2}$. Notice here that one of the active neutrinos is massless, since the neutrino mass matrix is rank 2.
Satisfying the neutrino oscillation data is rather easy task due to $O$, and all we should take care is the constraints of lepton flavor violations via $y_{L''L}$. The most stringent constraint arises from the process of $\mu\to e \gamma$.
To evade this constraint, we will take $y_{L''L}\lesssim{\cal O}(0.01)$, when the mediating mass scales inside the neutrino loop are ${\cal O}(100-1000)$ GeV. In this case, we cannot explain the sizable muon $g-2$, but we have another source that arises from the term $y_{e''e}$.
Thus we will concentrate on this term in the muon $g-2$ as well as LFVs below.
%%%%%%%%%

\subsection{LFV and muon $g-2$}

 {\it Muon $g-2$}:
The muon anomalous magnetic moment($\Delta a_\mu$) has been observed and its discrepancy is estimated by~\cite{Hagiwara:2011af}
\begin{align}
\Delta a_\mu=(26.1\pm8.0)\times10^{-10}.\label{eq:exp-g2mu}
\end{align}
%%%
Our $\Delta a_\mu$ is induced by interaction with $y_{e''e}$ coupling as explained above, and its form is computed as
\begin{align}
\Delta a_{\mu}&\approx \frac{2m_\mu^2}{(4\pi)^2} |(y_{e''e})_{12}|^2 F_{II}(m_S,M_{e''}),\\
%%%
F_{II}(m_a,m_b)&\equiv\frac{2 m_a^6+3m_a^4m_b^2-6m_a^2m_b^4+m_b^6+12m_a^4m_b^2\ln\left[\frac{m_b}{m_a}\right]}{12(m_a^2-m_b^2)^4},
\end{align}
where $m_S\equiv m_{S_{R_2}}\approx m_{S_{I_2}}$.
%%%

\begin{figure}[t]
\begin{center}
\includegraphics[width=70mm]{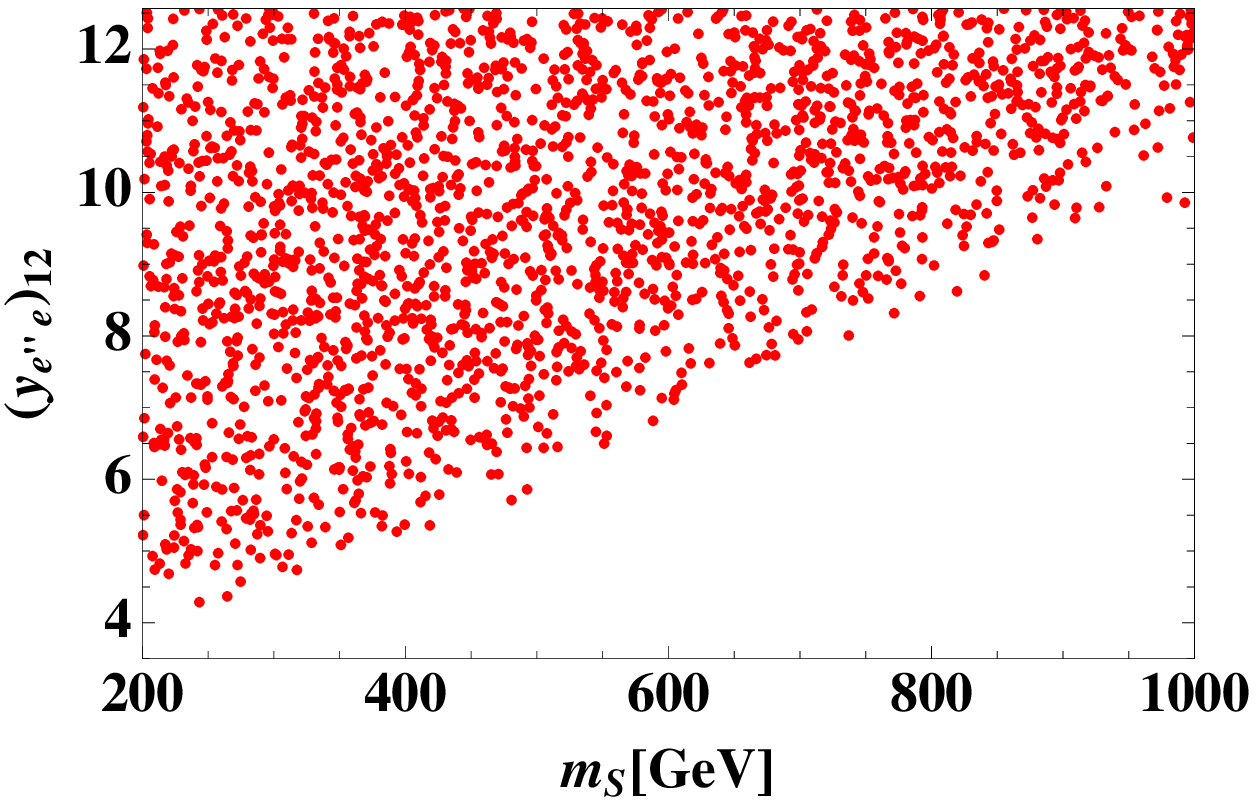} 
\qquad
\includegraphics[width=70mm]{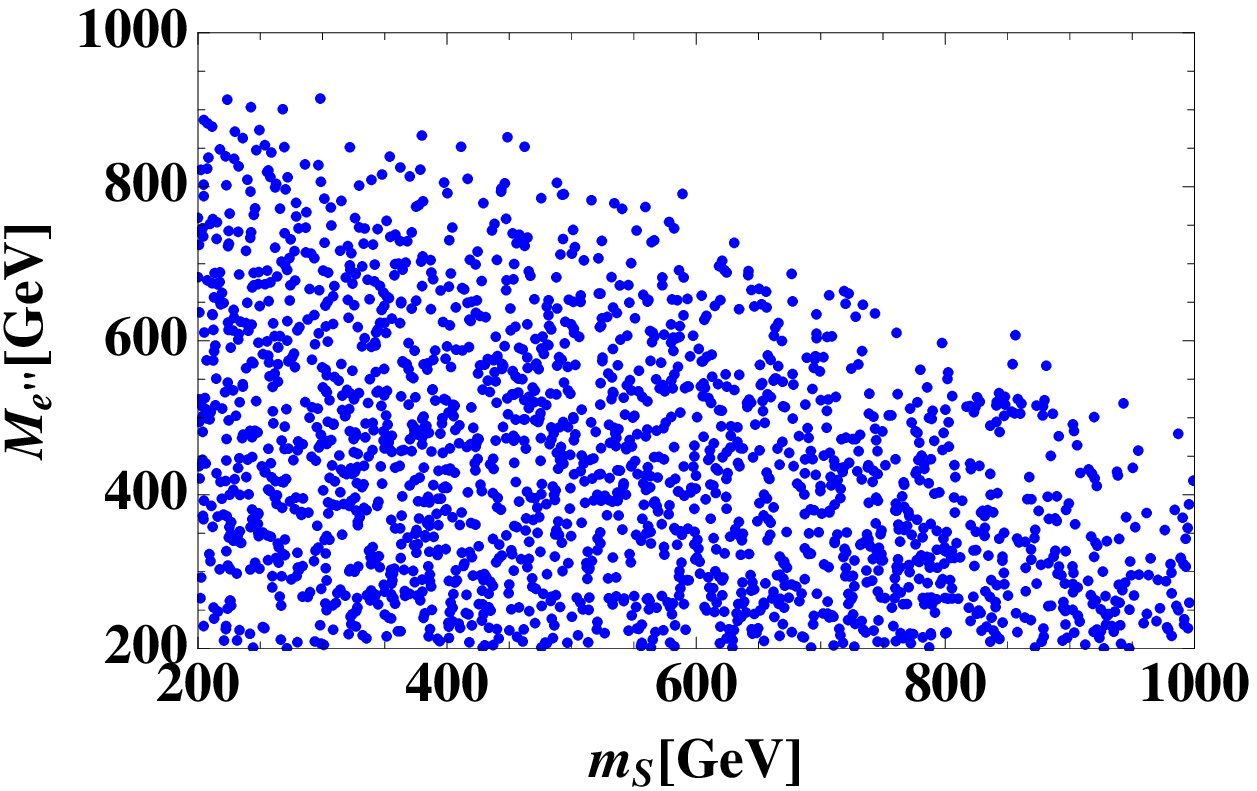}
\caption{Each of the left-side figure and right-side one represents the allowed points in terms of $m_S-(y_{e''e})_{12}$ and  $m_S-M_{e''e}$.} 
  \label{fig:g2mu}
\end{center}\end{figure}

{\it Lepton flavor violations (LFVs)}: LFV processes of $\ell \to \ell' \gamma$ are arisen from the same term as the $(g-2)_\mu$, and their forms are given by
\begin{align}
BR(\ell_a\to \ell_b \gamma)
&\approx\frac{48\pi^3C_{ab} \alpha_{em}}{(4\pi)^4G_F^2}
\left|\sum_{i=1}^3(y_{e''e}^\dag)_{b 1} (y_{e''e})_{1a} F_{lfv}(m_S, M_{e''})\right|^2,
\end{align}
where $\alpha_{em}\approx1/137$ is the fine-structure constant, $G_F\approx1.17\times10^{-5}$ GeV$^{-2}$ is the Fermi constant,
and $C_{21}\approx1$, $C_{31}\approx 0.1784$, $C_{32}\approx0.1736$. 
Experimental upper bounds are given by~\cite{TheMEG:2016wtm, Adam:2013mnn}: 
\[{\rm BR}(\mu\to e \gamma)\lesssim 4.2\times 10^{-13},\ 
{\rm BR}(\tau\to e \gamma)\lesssim 3.3\times 10^{-8},\ 
{\rm BR}(\tau\to \mu \gamma)\lesssim 4.4\times 10^{-13},
\]
where we define $\ell_1\equiv e$,  $\ell_2\equiv \mu$, and  $\ell_3\equiv \tau$. 
Comparing to the forms between muon $g-2$ and LFVs, one finds that putting the condition $(y_{e''e})_{11},(y_{e''e})_{13}<<(y_{e''e})_{12}$ 
provides the sizable muon $g-2$ without conflict of the constraints of LFVs. Under the condition,
we shows the allowed region to satisfy the sizable muon $g-2$ in fig.~\ref{fig:g2mu}. 
Each of the left-side figure and right-side one represents the allowed points in terms of $m_S-(y_{e''e})_{12}$ and  $m_S-M_{e''}$, where perturbative limit is set to be $y_{e''e}\lesssim4\pi$.
It suggests that rather large Yukawa coupling is required, while the mass ranges can widely be taken.

%%%%%%%%%%%%%%%%%%%
%%%%%%%%%%%%%%%%%%%

\subsection{ Dark matter} 
First of all, although we could have  the lightest mass of inert bosonic($s_{R{I}}$) or fermionc($\psi'_1$) DM candidate, the bosonic one requires rather small Yukawa coupling to satisfy the correct relic density $h^2\Omega\approx0.12$~\cite{Ade:2013zuv}, which conflicts with the sizable muon $g-2$ that demands the order one Yukawa coupling~\cite{Chiang:2017tai}.
% cannot  explain both of sizable muon $g-2$ and relic density of DM simultaneously  in our scenario~\cite{Chiang:2017tai}.
Thus we will focus on the lightest fermion $\psi'_1$ as the DM candidate, defining $\psi'_1 \equiv X$ and $M_{\psi'_1}\equiv M_X$.
%%%
{\it Furthermore we suppose that vertices $H_{1(2)}-X-X$ are tiny enough to avoid the lower bound of direct detection searches such as LUX~\cite{Akerib:2016vxi} and XENON1t~\cite{Aprile:2017iyp}.
It implies that Higgs portal contributions to the relic density of DM can be  negligible as well as Yukawa contributions via $y_{L''L}$. 
Thus DM annihilation processes are dominated by the gauge interaction with $Z'$.} 

\begin{figure}[t]
\begin{center}
\includegraphics[width=70mm]{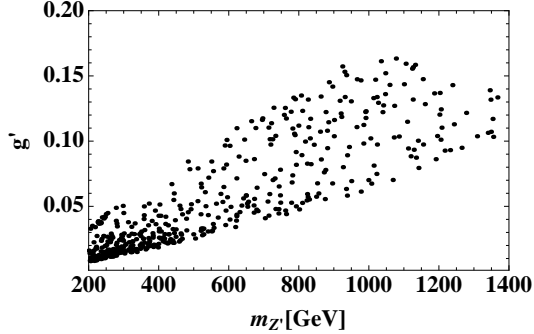}
\caption{The allowed points to satisfy the relic density of DM; $0.1\lesssim \Omega h^2\lesssim 0.15$, in terms of $m_{Z'}$ GeV and  $g'$. } 
  \label{fig:mZp-gp}
\end{center}\end{figure}

{\it Relic density}: Under the above set up, let us write down the valid Lagrangian in basis of mass eigenstate to contribute to the relic density as
\begin{align}
-{\cal L}&=
%%%
\frac{g'}{2}\left[-1+2 (U^*)_{14} (U^T)_{41}\right] \bar X\gamma^\mu \gamma^5 X Z'_\mu
%\nn\\&
-g'  \bar \ell_a \gamma^\mu \ell_a Z'_\mu - g'  \bar \nu_a \gamma^\mu P_L \nu_a Z'_\mu,
\end{align}
where we have used the unitary condition; $U^\dag U=1$ in the first term in the right side of above equation, $g'$ is $U(1)_L$ gauge coupling, and $a=1-3$.
Then the squared amplitude for the process $X\bar X\to \ell_a\bar\ell_a(\nu_a\bar \nu_a)$ via  $s$-channel is given by
\begin{align}
|\bar {\cal M} (X\bar X\to{\rm leptons})|^2 & \approx 
12 \left|\frac{g'^2\left[-1+2 (U^*)_{14} (U^T)_{41}\right] }{s-m_{Z'}^2 + i m_{Z'} \Gamma_{Z'}}\right|^2 G(M_X, m_{Z'}, \{p_i,k_i\}), 
\end{align}
\begin{align}
G(M_X, m_{Z'}, \{p_i,k_i\}) & \equiv  (p_1\cdot k_1)(p_2\cdot k_2)+(p_1\cdot k_2)(p_2\cdot k_1) +(M_X^2-p_1\cdot p_2)(k_1\cdot k_2-s) \nonumber \\
& -\frac{s}2(p_1\cdot p_2) -\left(\frac{s^2}{8m_{Z'}^2}\right) \left[(p_1\cdot k_1) +(p_2\cdot k_2)+(p_1\cdot k_2)+(p_2\cdot k_1)-s \right],
\end{align}
where $p_{1,2}(k_{1,2})$ denote initial(final) state momentum, $\Gamma_{Z'}$ is the total decay width of $Z'$, and masses of leptons are supposed to be massless. The inner products of momentum such as $(p_1\cdot p_2)$ are given in Ref.~\cite{Cheung:2016ypw}.
The decay width of $Z'$ is given by
\begin{equation}
\Gamma_{Z'} = \frac{g'^2m_{Z'}}{12 \pi} \sum_{f} N_c^{f} C_f (Q_{L}^{f})^2 \left( 1 + \frac{2 m_f^2}{m_{Z'}^2} \right) \sqrt{1- \frac{4 m_f^2}{m_{Z'}^2}},
\end{equation}
where we assume $Z'$ decays into only SM fermions $f$, $N_c^f$ is color factor, and $C_f = 1/2$ for neutrino while $C_f =1$ for the other fermions.
%Note also that $Z'$ mass is given by $m_{Z'} = g_{BL}v'$.
%%%
The relic density of DM is then given by~\cite{Edsjo:1997bg}
\begin{align}
&\Omega h^2
\approx 
\frac{1.07\times10^9}{\sqrt{g_*(x_f)}M_{Pl} J(x_f)[{\rm GeV}]},
\label{eq:relic-deff}
\end{align}
where $g^*(x_f\approx25)$ is the degrees of freedom for relativistic particles at temperature $T_f = M_X/x_f$, $M_{Pl}\approx 1.22\times 10^{19}$ GeV,
and $J(x_f) (\equiv \int_{x_f}^\infty dx \frac{\langle \sigma v_{\rm rel}\rangle}{x^2})$ is given by~\cite{Nishiwaki:2015iqa}
\begin{align}
J(x_f)&=\int_{x_f}^\infty dx\left[ \frac{\int_{4M_X^2}^\infty ds\sqrt{s-4 M_X^2} W(s) K_1\left(\frac{\sqrt{s}}{M_X} x\right)}{16  M_X^5 x [K_2(x)]^2}\right],\\ 
%%%%%%%%%%%
W(s)
\approx &\frac{1}{16\pi} \sum_a
\int_0^\pi \sin\theta
|\bar {\cal M} (X\bar X\to{\rm leptons})|^2,
%4 |\lambda_{SShh}|^2,
\label{eq:relic-deff}
\end{align}
where we implicitly impose the kinematical constraint above.
Then we numerically calculate relic density to search for the parameter region which can fit the observed data.
{The input parameters are randomly chosen in the ranges of
\begin{align}
& \{m_{N_1}, m_{N_2}, m_{N_3}, M'_{11}, M'_{12}, M'_{13}, m_{N'} \} \in [0.1, 10] \ {\rm TeV}, \nn \\
& m_{Z'} \in [M_X, 3 M_X], \quad g' \in [10^{-3}, 1],
\end{align}
where we also impose LEP constraint Eq.~(\ref{eq:lep}) and generate 5 million sampling points.}
We show the allowed region in terms of $m_{Z'}$ [GeV] and  $g'$ satisfying the correct relic density in fig.~\ref{fig:mZp-gp}
where we relax the allowed range of relic density of DM; $0.1\lesssim \Omega h^2\lesssim 0.15$, instead of the exact value.
We also find that the mass of DM is at around the half of  $m_{Z'}$, therefore we have a pole solutions.
Thus one finds that $M_X\lesssim 700$ GeV from this figure that is kinematically consistent with the analysis of muon $g-2$,
since the relation $700$ GeV $\lesssim (m_S, M_{e''})$ can be satisfied.

%%%%%%%%%%%%%%%%%%%
%%%%%%%%%%%%%%%%%%%

\subsection{ Colliders of $Z'$ boson} 

\begin{figure}[t]
\begin{center}
\includegraphics[width=70mm]{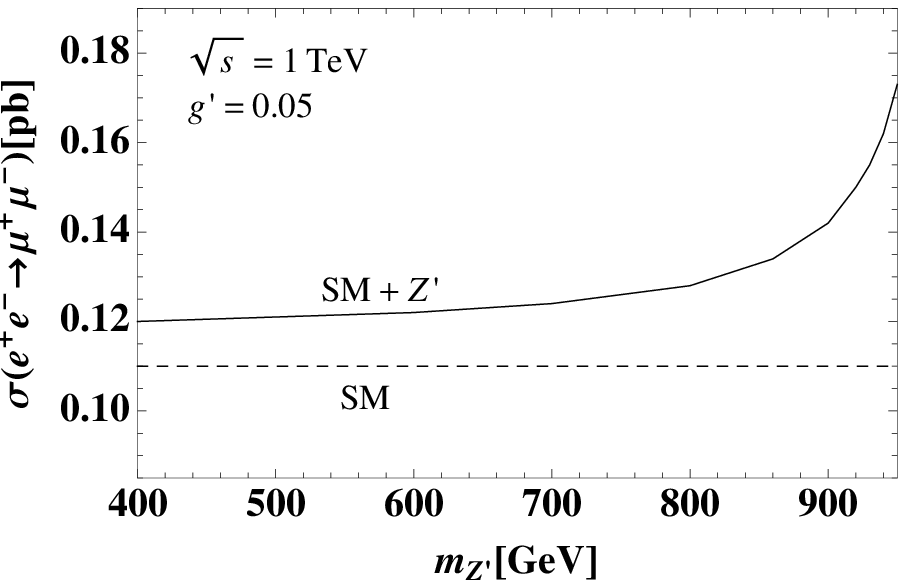}
\includegraphics[width=70mm]{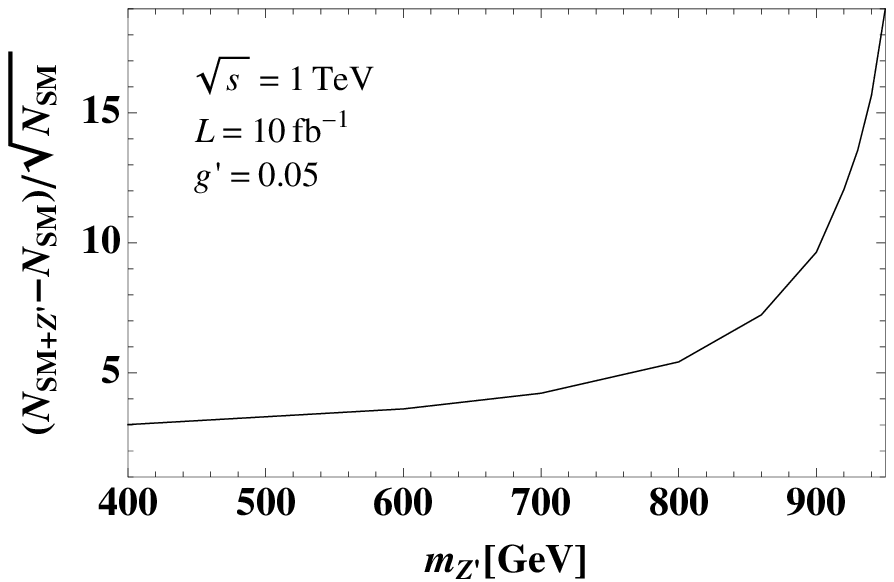}
\caption{Left: The cross section of $e^+ e^- \to \mu^+ \mu^-$ as a function of $m_{Z'}$ with $\sqrt{s} = 1$ TeV where SM prediction is also shown. Right: The ratio of number of signal events and square root of number of SM events as a function of $m_{Z'}$.  } 
  \label{fig:CX}
\end{center}\end{figure}

In our model, it is difficult to produce the $Z'$ boson directly at the LHC since it does not couples to quarks.
Then we discuss the signature of $Z'$ at the International Linear Collider (ILC) where $Z'$ can be produced by $e^+ e^- \to Z'$ process.
As a signature of $Z'$ we consider $e^+ e^- \to \mu^+ \mu^-$ process including both SM and $Z'$ contributions.
Here the production cross section for $e^+ e^- \to \mu^+ \mu^-$ is estimated using {\it CalcHEP}~\cite{Belyaev:2012qa} implementing relevant gauge interactions.
The left plot of Fig.~\ref{fig:CX} shows the cross section for $\sqrt{s}=1$ TeV as a function of $m_{Z'}$ where we apply $g' = 0.05$ as a reference value.
Note that the cross section becomes larger when the $Z'$ mass is close to 1 TeV due to the resonance effect.
We also estimate the significance of $Z'$ signature by $(N_{SM+Z'}-N_{SM})/\sqrt{N_{SM}}$ where $N_{SM}$ and $N_{SM+Z'}$ are number of events for only SM and SM with $Z'$ contribution respectively. Thus sizable significance is expected with integrated luminosity of 10 fb$^{-1}$ and our $Z'$ can be tested at the ILC. 

%%%%%%%%%%%%%%%%%%%
%%%%%%%%%%%%%%%%%%%

\section{Conclusion}
We have studied a radiative neutrino mass model in which lepton number symmetry is gauged adding $Z_2$ odd exotic leptons to cancel the gauge anomalies.
The active neutrino masses are generated at two-loop level where the exotic leptons propagate inside a loop diagram.
In addition, the lightest $Z_2$ odd neutral particle can be a good DM candidate.

We have formulated active neutrino mass matrix, partial decay width of LFV process $\ell \to \ell' \gamma$, muon $g-2$ and relic density of DM.
The muon $g-2$ can be as large as the observed value when we choose sizable Yukawa couplings and masses of exotic particles which is less than TeV scale. 
Relic density of our DM candidate is dominantly determined by the gauge interaction associated with $Z'$ and observed value can be obtained.
We have also discussed collider physics of $Z'$ which can be produced by $e^+e^-$ collision realized at ILC. 
Sizable significance of $Z'$ signal is expected for the parameter region which can explain the relic density of DM.

%\newpage
%%%%%%%%%%%%%%%%%%%%%%%%%%%%%%%%%%%
%\hspace{0.2cm} {\bf Acknowledgments}
%\section*{Acknowledgments}:
%\vspace{0.5cm}
\section*{Acknowledgments}
\vspace{0.5cm}
H. O. is sincerely grateful for the KIAS member and all around.
%%%%%%%%%%%%%%%%%%%%%%%%%%%%%%%%%%%
%%%%%%%%%%%%%%%%%%%%%%%%%%%%%%%%%%%

\end{document}